\documentclass[aps,twocolumn,floats,prl,nofootinbib,superscriptaddress,10pt]{revtex4-1}

\usepackage[utf8]{inputenc}

\usepackage[dvips]{graphicx} %
\usepackage{graphicx,amsmath,amsfonts,amssymb,slashed}
\usepackage{bbold,wasysym}
\usepackage{graphicx}

\usepackage[usenames,dvipsnames]{xcolor} 

\usepackage{soul}

\definecolor{RedWine}{rgb}{0.743,0,0}
\definecolor{RoyalBlue}{rgb}{0.25,.41,.88}

\setstcolor{Blue}

\newcommand{\VEV}[1]{\left<{1}\right>}

\begin{document}

\title{Did LIGO detect dark matter?}

\author{Simeon Bird}
\email{E-mail: sbird4@jhu.edu}
\noaffiliation
\author{Ilias Cholis}
\noaffiliation
\author{Julian B.\ Mu\~noz}
\noaffiliation
\author{Yacine Ali-Ha\"imoud}
\noaffiliation
\author{Marc Kamionkowski}
\noaffiliation
\author{Ely D.\ Kovetz}
\noaffiliation
\author{Alvise Raccanelli}
\noaffiliation
\author{Adam G.\ Riess}
\affiliation{Department of Physics and Astronomy, Johns Hopkins
     University, 3400 N.\ Charles St., Baltimore, MD 21218, USA}

\begin{abstract}
We consider the possibility that the black-hole (BH) binary detected
by LIGO may be a signature of dark matter.
Interestingly enough, there remains a window for masses $20\,M_\odot
\lesssim M_{\rm bh} \lesssim 100\, M_\odot$ where primordial
black holes (PBHs) may constitute the dark matter.  If two
BHs in a galactic halo pass sufficiently close, they radiate
enough energy in gravitational waves to become gravitationally
bound.  The bound BHs will rapidly spiral inward due to emission of
gravitational radiation and ultimately merge.  Uncertainties in
the rate for such events arise from our imprecise knowledge of
the phase-space structure of galactic halos on the smallest
scales.  Still, reasonable estimates span a range that overlaps the
$2-53$~Gpc$^{-3}$~yr$^{-1}$ rate estimated from GW150914, thus
raising the possibility that LIGO has detected PBH dark matter.
PBH mergers are likely to be distributed spatially more like
dark matter than luminous matter and have no optical nor
neutrino counterparts. They may be distinguished from mergers of BHs
from more traditional astrophysical sources through the observed mass
spectrum, their high ellipticities, or their stochastic gravitational wave background. 
Next generation experiments will be invaluable in performing these tests.
\end{abstract}

\maketitle

The nature of the dark matter (DM) is one of the most longstanding
and puzzling questions in physics.  Cosmological
measurements have now determined with exquisite precision the
abundance of DM \cite{Hinshaw:2012aka,Ade:2015xua}, and
from both observations and numerical simulations we know quite a bit 
about its distribution in Galactic halos. Still, the nature of the DM remains a
mystery.  Given the efficacy with which weakly-interacting
massive particles---for many
years the favored particle-theory explanation---have eluded
detection, it may be warranted to consider other possibilities
for DM.  Primordial black holes (PBHs) are one such
possibility \cite{Carr:1975qj, Carr:1974nx, Meszaros:1974tb, Clesse:2015wea}.

Here we consider whether the two $\sim30\,M_\odot$ black holes
detected by LIGO \cite{Abbott:2016blz} could plausibly be PBHs.
There is a window for PBHs to be DM if the BH mass is in
the range $20\,M_\odot \lesssim M \lesssim
100\,M_\odot$ \cite{Carr:2009jm,Monroy-Rodriguez:2014ula}.
Lower masses are excluded by microlensing surveys \cite{Allsman:2000kg, Tisserand:2006zx,Wyrzykowski:2011tr}.
Higher masses would disrupt wide binaries \cite{Yoo:2003fr,Quinn:2009zg,Monroy-Rodriguez:2014ula}.
It has been argued that PBHs in this mass range are excluded by CMB constraints \cite{Ricotti:2007au, Ricotti:2007jk}. 
However, these constraints require modeling of 
several complex physical processes, including the accretion of gas onto a moving 
BH, the conversion of the accreted mass to a luminosity, the self-consistent 
feedback of the BH radiation on the accretion process, and the deposition of the 
radiated energy as heat in the photon-baryon plasma.  A significant (and 
difficult to quantify) uncertainty should therefore be associated with this 
upper limit \cite{AliHaimoud:2016}, and it seems worthwhile to examine whether PBHs 
in this mass range could have other observational consequences.

In this {\sl Letter}, we show that if DM consists
of $\sim 30~M_\odot$ BHs, then the rate for mergers of such
PBHs falls within the merger rate inferred from GW150914.
In any galactic halo, there is a chance two BHs
will undergo a hard scatter, lose energy to a
soft gravitational wave (GW) burst and become gravitationally
bound.  This BH binary will merge via emission of
GWs in less than a Hubble time.\footnote{In our analysis, PBH binaries 
are formed inside halos at $z=0$. Ref.~\cite{Nakamura:1997sm}
considered instead binaries which form at early times and merge over a Hubble time.}
Below we first estimate roughly the rate of such mergers and then present
the results of more detailed calculations.  We
discuss uncertainties in the calculation and some possible
ways to distinguish PBHs from BH binaries from more traditional
astrophysical sources. 

Consider two PBHs approaching each other 
on a hyperbolic orbit with some impact parameter and relative velocity $v_\mathrm{pbh}$.
As the PBHs near each other, they produce a time-varying quadrupole moment 
and thus GW emission.
The PBH pair becomes gravitationally bound if the GW
emission exceeds the initial kinetic energy. The cross
section for this process is \cite{Quinlan:1989,Mouri:2002mc},
\begin{eqnarray}
     \sigma &=& \pi \left( \frac{85\, \pi}{3}
     \right)^{2/7} R_{s}^2 \left(\frac{v_\mathrm{pbh}}{c}\right)^{-18/7}
     \nonumber \\ &=&
     1.37 \times 10^{-14}\,M_{30}^2\, v_{\mathrm{pbh}-200}^{-18/7}\,{\rm
     pc}^2,
\label{eqn:crosssection}
\end{eqnarray}
where $M_\mathrm{pbh}$ is the PBH
mass, and $M_{30}$ the PBH mass in units of $30\,M_\odot$, 
$R_{s}= 2 G M_\mathrm{pbh}/c^2$ is its Schwarzschild
radius,  $v_\mathrm{pbh}$ is the relative velocity of two PBHs,
and $v_{\mathrm{pbh}-200}$ is this velocity in units of $200$~km~sec$^{-1}$.

We begin with a rough but simple and illustrative
estimate of the rate per unit volume of such mergers.  Suppose
that all DM in the Universe resided
in Milky-Way like halos of mass $M=M_{12} \,
10^{12}\,M_\odot$ and uniform mass density $\rho= 0.002\, \rho_{0.002}\,
M_\odot$~pc$^{-3}$ with $\rho_{0.002}\sim1$.  
Assuming a uniform-density halo of volume $V = M/\rho$,
the rate of mergers per halo would be
\begin{eqnarray} 
     N &\simeq & (1/2) V (\rho/M_{\rm pbh})^2 \sigma v
     \nonumber \\
     & \simeq& 3.10 \times 10^{-12}\, M_{12} \, \rho_{0.002}\,
     v_{\mathrm{pbh}-200}^{-11/7}\, {\rm yr}^{-1}\,.
\label{eqn:rateperhalo}
\end{eqnarray}
The relative velocity $v_{\mathrm{pbh}-200}$ is specified 
by a characteristic halo velocity. The mean cosmic DM mass density
is $\rho_{\rm dm} \simeq 3.6\times 10^{10}\,
M_\odot$~Mpc$^{-3}$, and so the spatial density of halos is
$n\simeq 0.036\,M_{12}^{-1}$~Mpc$^{-3}$.  The rate per
unit comoving volume in the Universe is thus
\begin{equation}
     \Gamma \simeq 1.1 \times 10^{-4} \, \rho_{0.002} \,
     v_{\mathrm{pbh}-200}^{-11/7} \, {\rm Gpc}^{-3}\, {\rm yr}^{-1}.
\label{eqn:volumerate}
\end{equation}
The normalized halo mass $M_{12}$ drops out, as it should. The merger
rate per unit volume also does not depend on the PBH
mass, as the capture cross section scales like $M_\mathrm{pbh}^2$.

This rate is small compared with the
$2-53$~Gpc$^{-3}$~yr$^{-1}$ estimated by LIGO for a population of
$\sim30\,M_\odot - 30 \, M_\odot$ mergers \cite{Abbott:2016nhf}, 
but it is a very conservative estimate.  As
Eq.~(\ref{eqn:volumerate}) indicates, the merger rate is
higher in higher-density regions and in regions of lower DM
velocity dispersion.  The DM in Milky-Way like halos is known
from simulations \cite{Moore:1999nt} and analytic models
\cite{Kamionkowski:2008vw} to have substructure, regions of
higher density and lower velocity dispersion.  DM halos also have a
broad mass spectrum, extending to very low masses
where the densities can become far higher, and velocity
dispersion far lower, than in the Milky Way.  To get a very rough
estimate of the conceivable increase in the PBH merger rate due
to these smaller-scale structures, we can replace $\rho$ and $v$
in Eq.~(\ref{eqn:volumerate}) by the values they would have had
in the earliest generation of collapsed objects, where the
DM densities were largest and velocity dispersions
smallest.  If the primordial power spectrum
is nearly scale invariant, then gravitationally bound halos of
mass $M_c \sim 500~M_\odot$, for example, will form at redshift 
$z_c\simeq 28 - \log_{10}(M_c/500\,M_\odot)$.  These objects will have virial
velocities $v\simeq 0.2$~km~sec$^{-1}$ and densities $\rho \simeq
0.24~M_\odot$~pc$^{-3}$ \cite{Kamionkowski:2010mi}.  Using
these values in Eq.~(\ref{eqn:volumerate}) increases the merger
rate per unit volume to 
\begin{equation}
     \Gamma \simeq 700\, {\rm Gpc}^{-3}\,{\rm yr}^{-1}.
\label{eqn:bigvolumerate}
\end{equation}
This would be the merger rate if \emph{all} the DM resided in the smallest haloes.
Clearly, this is not true by the present day; substructures are at 
least partially stripped as they merge to form larger objects, and so
Eq.~(\ref{eqn:bigvolumerate}) should be viewed as a conservative upper limit.

Having demonstrated that rough estimates contain the merger-rate
range $2-53$~Gpc$^{-3}$~yr$^{-1}$ suggested by LIGO, we now
turn to more careful estimates of the PBH merger rate.
As Eq.~(\ref{eqn:volumerate}) suggests, the merger rate will
depend on a density-weighted average, over the entire cosmic
DM distribution, of $\rho_{0.002} v_{\mathrm{pbh}-200}^{-11/7}$.  
To perform this average, we will (a) assume that DM
is distributed within galactic halos with a Navarro-Frenk-White (NFW)
profile \cite{Navarro:1995iw} with concentration parameters
inferred from simulations; and (b) try several halo mass
functions taken from the literature for the distribution of halos.

The PBH merger rate $\mathcal{R}$ within each halo can be
computed using
\begin{equation}
      \mathcal{R} =  4 \pi \int^{R_\mathrm{vir}}_0 r^2
      \frac{1}{2}\left(\frac{\rho_{\mathrm{nfw}} (r) }{
      M_\mathrm{pbh}}\right)^2 \left<\sigma
      v_\mathrm{pbh}\right> \,dr
 \label{eq:rate}
\end{equation}
where $\rho_{\mathrm{nfw}}(r) = \rho_s \left[
(r/R_s)(1+r/R_s)^2 \right]^{-1}$ is the NFW
density profile with characteristic radius $r_s$ and characteristic density $\rho_s$. 
$R_\mathrm{vir}$ is the virial radius at which the NFW profile reaches a value $200$ times 
the comoving mean cosmic density and is cutoff.
The angle brackets denote an average over the PBH
relative velocity distribution in the halo. The merger cross
section $\sigma$ is given by Eq.~(\ref{eqn:crosssection}). 
We define the concentration parameter $C = R_\mathrm{vir}/R_s$. 
To determine the profile of each halo, we require $C$ as a function of halo mass $M$. 
We will use the concentration-mass relations fit to DM N-body simulations 
by both Ref.~\cite{Prada:2011jf} and Ref.~\cite{Ludlow:2016ifl}.


We now turn to the average of the cross section times relative
velocity. The one-dimensional velocity dispersion of a halo is
defined in terms of the escape velocity at radius $R_\mathrm{max} =
2.1626\, R_s$, the radius of the maximum circular velocity of
the halo. i.e.,
\begin{equation}
      v_\mathrm{dm} = \sqrt{\frac{G M(r < r_\mathrm{max})}{r_\mathrm{max}}}
  = \frac{v_\mathrm{vir}}{\sqrt{2}} \sqrt{ \frac{C}{C_m}
  \frac{g(C_m)}{g(C)}},
\end{equation}
where $g(C) = ln(1+C) - C/(1+C)$, and $C_m = 2.1626 =
R_\mathrm{max} / R_s$. We approximate the relative velocity distribution
of PBHs  within a halo as a Maxwell-Boltzmann (MB) distribution with
a cutoff at the virial velocity. i.e.,
\begin{equation}
     P(v_\mathrm{pbh}) = F_0 \left[\exp{\left(
      -\frac{v^2_\mathrm{pbh}}{v_\mathrm{dm}^2}\right)}  - \exp{\left(
       -\frac{v^2_\mathrm{vir}}{v_\mathrm{dm}^2}\right)}\right]\,,
\label{eqn:veldisp}
\end{equation}
where $F_0$ is chosen so that $4\pi \int^{v_\mathrm{vir}}_0 P(v) v^2 dv = 1$.
This model provides a reasonable match to N-body simulations, at
least for the velocities substantially less than than the virial
velocity which dominate the merger rate (e.g.,
Ref.~\cite{Mao:2012hf}). Since the cross-section is
independent of radius, we can integrate 
the NFW profile to find the merger rate in any halo:
\begin{align}
     \mathcal{R} =  \left(\frac{85 \pi}{12\sqrt{2}}\right)^{2/7}
     \frac{9 G^2 M_\mathrm{vir}^2}{c R_s^3}\left(1-\frac{1}{(1+C)^3}\right)
     \frac{D(v_\mathrm{dm})}{g(C)^2}\,,
     \label{eqn:halomergerrate}
\end{align}
where
\begin{equation}
     D(v_\mathrm{dm}) = \int^{v_\mathrm{vir}}_0 P(v, v_\mathrm{dm})
     \left(\frac{2 v}{c}\right)^{3/7} dv,\
\end{equation}
comes from Eq.~(\ref{eqn:veldisp}).

Eq.~(\ref{eqn:crosssection}) gives the cross section for two
PBHs to form a binary. However, if the binary is to
produce an observable GW signal, these
two PBHs must orbit and inspiral; a
direct collision, lacking an inspiral phase, is 
unlikely to be detectable by LIGO. 
This requirement imposes a minimum impact parameter of
roughly the Schwarzschild radius. The fraction of BHs
direct mergers is $\sim v^{2/7}$ and reaches a maximum
of $\sim 3\%$ for $v_\mathrm{pbh} = 2000$ km s$^{-1}$.  Thus, direct
mergers are negligible.  We also require that once the binary is
formed, the time until it merges (which can be obtained from
Ref.~\cite{O'Leary:2008xt}) is less than a Hubble time.  
The characteristic time it takes for a binary BH
to merge varies as a function of halo velocity dispersion. It can be
hours for $M_\mathrm{vir} \simeq 10^{12}\,M_{\odot}$
or kyrs for $M_\mathrm{vir} \simeq 10^{6}\, M_{\odot}$, and is thus
instantaneous on cosmological timescales.  Given the
small size of the binary, and rapid time to merger, we can neglect disruption
of the binary by a third PBH once formed.
BH binaries can also form through non-dissipative three-body encounters. 
The rate of these binary captures is non-negligible in small halos 
\cite{Quinlan:1989, Lee:1993}, but they generically lead 
to the formation of wide binaries that will not be able to harden and 
merge within a Hubble time. This formation mechanism 
should not affect our LIGO rates. The merger rate is therefore equal 
to the rate of binary BH formation, Eq.~(\ref{eqn:halomergerrate}).

\begin{figure}
\includegraphics[width=0.48\textwidth]{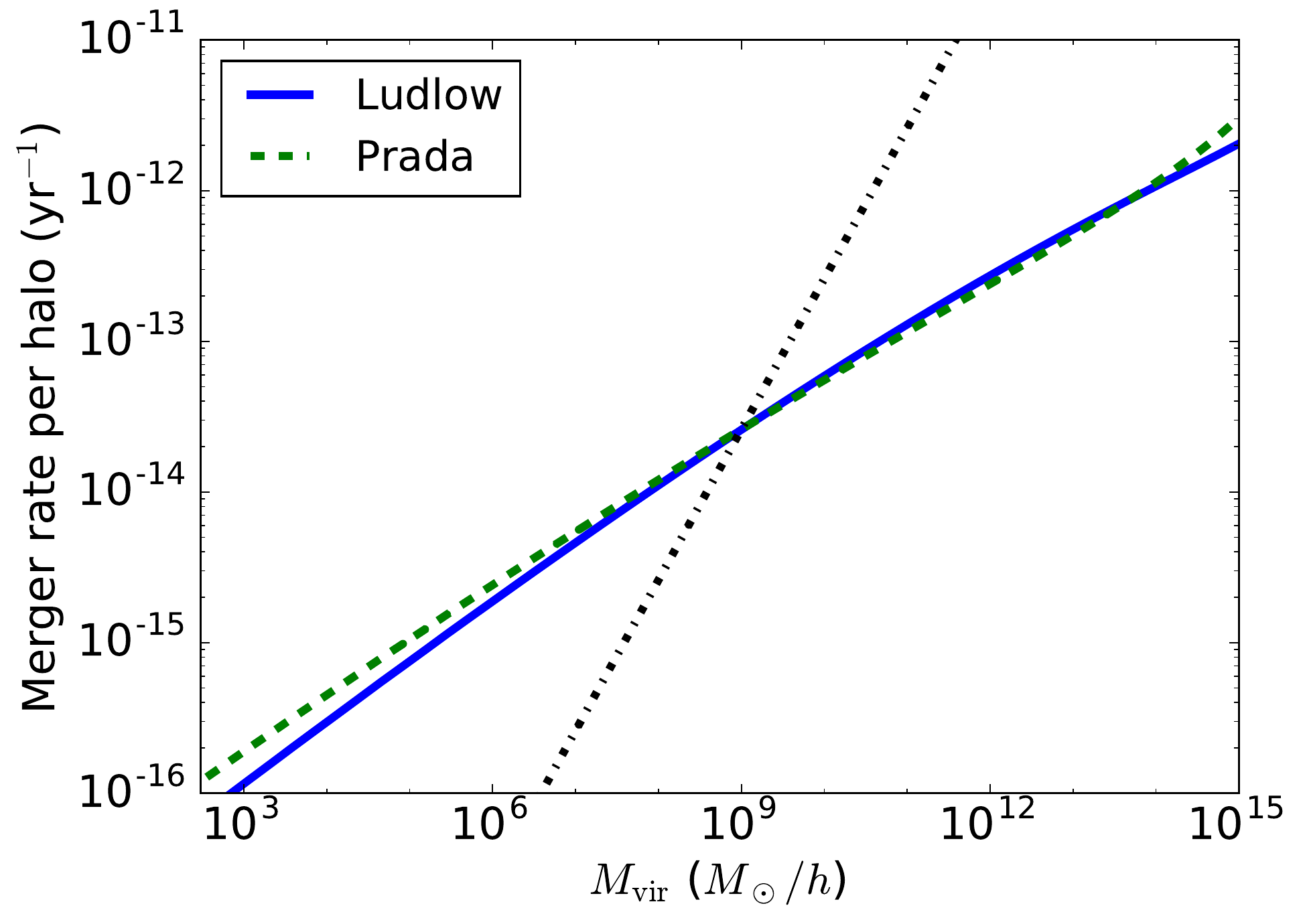}
\caption{The PBH merger rate per halo as a function of halo
      mass. The  solid line shows the trend assuming the
      concentration-mass relation from
      Ref.~\protect\cite{Ludlow:2016ifl}, and the dashed line
      that from Ref.~\protect\cite{Prada:2011jf}.  To guide the 
      eye, the dot-dashed line shows a constant BH merger rate per unit
      halo mass.}
 \label{fig:halomergerrate}
\end{figure}

Fig.~\ref{fig:halomergerrate} shows the contribution to the
merger rate, Eq.~(\ref{eqn:halomergerrate}), for two concentration-mass 
relations. As can be seen, both concentration-mass relations give similar results. 
An increase in halo mass produces an increased PBH merger
rate. However, less massive halos have a higher concentration
(since they are more likely to have virialized earlier), so that
the merger rate per unit mass increases significantly
as the halo mass is decreased.

To compute the expected LIGO event rate, we convolve the merger
rate $\mathcal{R}$ per halo with the mass function $dn/dM$.  Since the
redshifts ($z \lesssim 0.3$) detectable by LIGO are relatively low
we will neglect redshift evolution in the halo mass
function. The total merger rate per unit volume is then,
\begin{equation}
     \mathcal{V} = \int\, (dn/dM)(M)\, \mathcal{R}(M)\, dM.
\label{eqn:halovolumerate}
\end{equation}
Given the exponential falloff of $dn/dM$ at high masses, despite 
the increased merger rate per halo suggested in
Fig.~\ref{fig:halomergerrate}, the precise value of the upper
limit of the integrand does not affect the final result.

At the lower limit, discreteness in the DM particles
becomes important, and the NFW profile is no longer a good  
description of the halo profile.  Furthermore, the smallest
halos will evaporate due to periodic ejection of objects by
dynamical relaxation processes. The evaporation timescale is
\cite{Binney:1987}
\begin{equation}
     t_{\rm evap} \approx (14\,\mathcal{N}/\ln \mathcal{N})
     \left[R_\mathrm{vir}/(C \,v_\mathrm{dm}) \right],
\end{equation}
where $\mathcal{N}$ is the number of individual BHs in the halo,
and we assumed that the PBH mass is $30\, M_{\odot}$.  For a halo of
mass $400\, M_\odot$, the velocity dispersion is
$0.15$~km~sec$^{-1}$, and the evaporation timescale is $\sim 3$
Gyr.  In practice, during matter domination, halos which have already 
formed will grow continuously through mergers or accretion. Evaporation will 
thus be compensated by the addition of new material, and as halos grow
new halos will form from mergers of smaller objects.
However, during dark-energy domination at $z \lesssim 0.3$, $3$ Gyr ago, 
this process slows down.  Thus, we will neglect the signal from
halos with an evaporation timescale less than $3$ Gyr, 
corresponding to $M < 400\, M_\odot$. This is in any case $13$
PBHs, and close to the point where the NFW profile is no longer valid.

\begin{figure}
\includegraphics[width=0.48\textwidth]{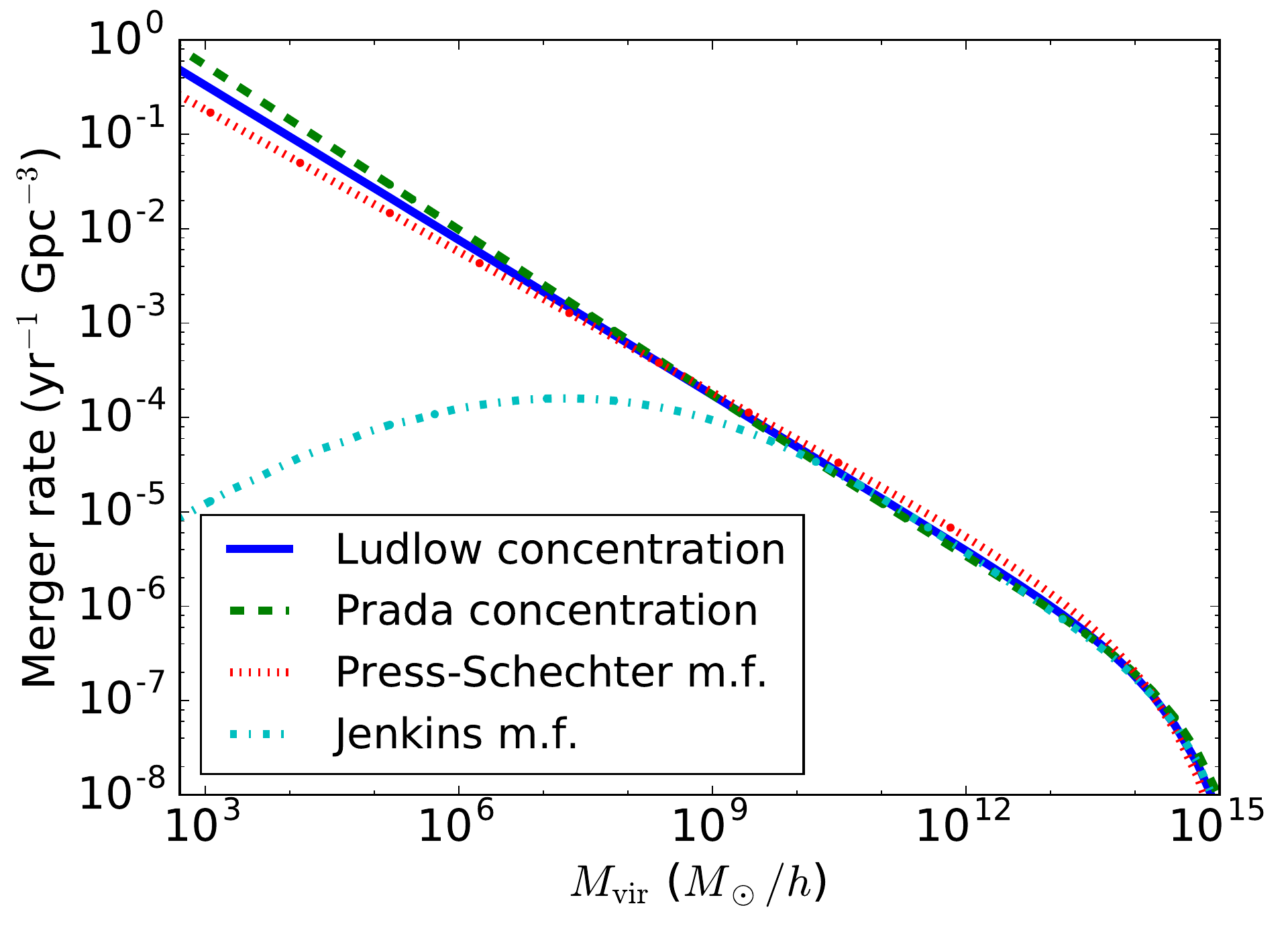}
\caption{The total PBH merger rate as a function of halo mass.
      Dashed and dotted lines show different prescriptions for
      the concentration-mass relation and halo mass function.}
\label{fig:volume}
 \label{fig:totalhalomergerrate}
\end{figure}

The halo mass function $dn/dM$ is computed using both semi-analytic
fits to N-body simulations and with analytic
approximations. Computing the merger rate in the small halos
discussed above requires us to extrapolate both the halo mass
function and the concentration-mass relation around six orders
of magnitude in mass beyond the smallest halos present in the calibration
simulations. High-resolution 
simulations of $10^{-4} M_\odot$ cold dark matter micro-halos 
\cite{Ishiyama:2010es,Ishiyama:2014uoa} suggest that our 
assumed concentration-mass relations underestimate the internal 
density of these halos, making our rates conservative.

The mass functions depend on the halo mass through the perturbation amplitude $\sigma(R_{\rm vir})$ 
at the virial radius $R_{\rm vir}$ of a given halo. Due to the scale invariance of
the window functions on small scales, $\sigma(R_{\rm vir})$ varies only by a
factor of two between $M_\mathrm{vir} = 10^9\,M_\odot$ and $10^3\, M_\odot$. Thus
the extrapolation in the mass function is less severe than it
looks.  We also note that the scale-invariant nature of the
initial conditions suggests that the shape of the halo mass
function should not evolve unduly until it reaches the scale of
the PBH mass, or evaporation cutoff. 

To quantify the uncertainty induced by the $dn/dM$
extrapolation, we obtained results with two different
mass functions: the classic analytic Press-Schechter 
calculation \cite{Press:1973iz} and one calibrated to numerical 
simulations from Tinker et al.~\cite{Tinker:2008ff}. The agreement
between the two small-scale behaviors suggests that extrapolating 
the mass functions is not as blind as it might otherwise seem.  
We also include a third mass function,
due to Jenkins et. al. \cite{Jenkins:2000bv}, that includes an artificial
small-scale mass cutoff at a halo mass $M_\mathrm{vir}\sim 10^6\,M_\odot$.
This cutoff is inserted to roughly model the mass
function arising if there is no power on scales smaller than
those currently probed observationally. We include it 
to provide a \emph{very} conservative lower limit to the
merger rate if, for some reason, small-scale power were
suppressed.  We do not, however, consider it likely that this
mass function accurately represents the distribution of halo masses 
in our Universe.

Fig.~\ref{fig:volume} shows the merger rate per logarithmic
interval in halo mass.  In all cases, halos with $M_\mathrm{vir} \lesssim
10^9\, M_\odot$ dominate the signal, due to the increase in
concentration and decrease in velocity dispersion with smaller
halo masses. The Tinker mass function, which asymptotes to a
constant number density for small masses, produces the most
mergers.  Press-Schechter has $\sim 50\%$ fewer events in small
halos, while the Jenkins mass function results in merger rates
nearly four orders of magnitude smaller (and in rough agreement
with Eq.~(\ref{eqn:volumerate})).

We integrate the curves in Fig.~\ref{fig:volume} to compute the total merger
rate $\mathcal{V}$.  All mass functions give a similar result, $\sim (3 \pm
1)\times 10^{-4}$~Gpc$^{-3}$~yr$^{-1}$, from halos of masses
$\gtrsim 10^{9}\, M_\odot$, representing for the Tinker and
Press-Schechter mass function a small fraction of the events.
When we include all halos with $M_\mathrm{vir} > 400 M_\odot$, the number of
events increases dramatically, and depends strongly on the
lower cutoff mass $M_c$ for the halo mass. 
Both the Press-Schechter and Tinker mass functions are for small
halos linear in the integrated perturbation amplitude $\propto
1/\sigma(R_{\rm vir})$ at the virial radius $R_{\rm vir}$
of the collapsing halo. In small halos, $1/\sigma(R_{\rm vir})$
is roughly constant.  Thus for a mass function ${\rm MF}(\sigma)$, we have
\begin{equation}
      (dn/dM) \sim (C \log \sigma/dM) \left[{\rm MF}(\sigma)/M_\mathrm{vir}
      \right] \sim M_\mathrm{vir}^{-2}.
\end{equation}
The concentration is also a function of $1/\sigma(R_{\rm vir})$
and it too becomes roughly constant for small masses. Assuming a
constant concentration,  the merger rate per halo scales as
$\mathcal{R} \sim M^{10/21}$.  Thus,
Eq.~(\ref{eqn:halovolumerate}) suggests that $\mathcal{V} \sim
M_c^{-11/21}$. This compares well to numerical differentiation
of Fig.~\ref{fig:volume}, which yields $\mathcal{V} \sim
M_c^{-0.51}$.

The integrated merger rate is thus
\begin{equation}
     \mathcal{V} = 2\, f (M_c/400\,M_\odot)^{-11/21}\,{\rm
     Gpc}^{-3}\, {\rm yr}^{-1},
\label{eqn:finalresult}
\end{equation}
with $f\simeq 1$ for the Tinker mass function, and $f
\simeq 0.6$ for the Press-Schechter mass function (the Jenkins mass
function results in an event rate $\mathcal{V} \simeq
0.02$~Gpc$^{-3}$~yr$^{-1}$, independent of $M_c \lesssim 10^6 M_\odot$).

A variety of astrophysical processes may alter the mass function
in some halos, especially within the dwarf galaxy range, $10^{9}
- 10^{10} M_\odot$.  However, halos with $M_\mathrm{vir} \lesssim 10^{9}\,
M_\odot$ are too small to form stars against the thermal
pressure of the ionized intergalactic medium
\cite{Efstathiou:1992zz} and are thus unlikely to be affected by
these astrophysical processes.  Inclusion of galactic
substructure, which our calculation neglects, should boost the
results.  However, since the event rate is dominated by the
smallest halos, which should have little substructure, 
we expect this to make negligible difference to our final result. 

There is also the issue of the NFW density profile assumed.  The
results are fairly insensitive to the detailed density profile
as long as the slope of the density profile varies no more
rapidly than $r^{-1}$ as $r\to0$.  For example, suppose we replace
the NFW profile with the Einasto profile \cite{Einasto:1965},
\begin{equation}
 \rho(R) = \rho_0 \exp\left(-\frac{2}{\alpha}\left[ \left(\frac{R}{R_\mathrm{s}}\right)^\alpha - 1 \right] \right)
\end{equation}
with $\alpha = 0.18$, which has a core as $r\to0$. The reduction
in the merger rate as $r\to0$ is more than compensated by an
increased merger rate at larger radii leading to a total merger
rate that is raised by $50\%$ relative to NFW, to 
$\sim 3$~Gpc$^{-3}$~yr$^{-1}$.  

Our assumption of an isotropic MB-like velocity distribution in the halo may also
underestimate the correct answer, as any other velocity
distribution would have lower entropy and thus larger averaged $v^{-11/7}$.  
Finally, the discreteness of PBH DM will provide some Poisson 
enhancement of power on $\sim 400\,M_\odot$ scales.  
More small-scale power would probably lead to an enhancement 
of the event rate beyond Eq.~(\ref{eqn:finalresult}).

The recent LIGO detection of two merging $\sim 30\, M_\odot$
black holes suggests a 90\% C.L.\ event rate
\cite{Abbott:2016nhf} of $2-53$~Gpc$^{-3}$~yr$^{-1}$ if all
mergers have the masses and emitted energy of GW150914.  
{\it It is interesting that---although there are theoretical uncertainties---our best
estimates of the merger rate for $30\,M_\odot$ PBHs, obtained
with canonical models for the DM distribution, fall in
the LIGO window}.  

The possibility that LIGO has seen DM thus cannot be immediately excluded.
Even if the predicted merger rates turn out, with more precise
treatments of the small-scale galactic phase-space distribution,
to be smaller, conservative lower estimates of the merger rate
for PBH DM suggests that the LIGO/VIRGO network should
see a considerable number of PBH mergers over its lifetime.

We have assumed a population of PBHs with the same mass.  
The basic results obtained here should, however,
remain unaltered if there is some small spread of PBH masses, as
expected from PBH-formation scenarios, around the nominal
value of $30\,M_\odot$.

PBH mergers may also be interesting for LIGO/VIRGO even if PBHs make
up only a fraction $f_\mathrm{pbh}$ of the DM, as 
implied by CMB limits from Refs.~\cite{Ricotti:2007jk, Ricotti:2007au} or 
the limits in Ref.~\cite{Carr:2009jm}.  
In this case, the number
density of PBHs will be reduced by $f_\mathrm{pbh}$. The 
cutoff mass will increase as $M_c \sim f^{-1}_\mathrm{pbh}$ if we
continue to require  $>13$ PBHs in each halo to avoid halo evaporation. 
The overall event rate will be 
$\mathcal{V} \sim 2 f_\mathrm{pbh}^{53/21}$~Gpc$^{-3}$~yr$^{-1}$.
Advanced LIGO will reach design sensitivity in 2019 \cite{Aasi:2013wya,TheLIGOScientific:2016zmo}, and
will probe $z <0.75$, an increase in volume to $\approx 50$ Gpc$^{3}$ (comoving).
Thus over the six planned years of aLIGO operation, while we should
expect to detect $\sim 600$ events with $f_\mathrm{pbh} = 1$,
we will expect at least one event if $f_\mathrm{pbh} > 0.1$.

Distinguishing whether any individual GW event, or
even some population of events, are from PBH DM or 
more traditional astrophysical sources will be daunting.  Still,
there are some prospects.  Most apparently, PBH mergers will be
distributed more like small-scale DM halos and are thus
less likely to be found in or near luminous galaxies than BH
mergers resulting from stellar evolution. Moreover,
PBH mergers are expected to have no electromagnetic/neutrino
counterparts whatsoever.  A DM component could conceivably
show up in the BH mass spectrum as an excess of events with BH
masses near $30\,M_\odot$ over a more broadly distributed mass
spectrum from astrophysical sources \cite[e.g.][]{Belczynski:2016obo}.

Since the binary is formed on a very elongated orbit, the GW
waveforms will initially have high ellipticity, exhibited by 
higher frequency harmonics in the GW signal \cite{O'Leary:2008xt}.  We have verified  
that the ellipticities become unobservably small by the time the inspiral enters the 
LIGO band, but they may be detectable in future experiments \cite{ellipticity}. 
Mutiply-lensed quasars \cite{Pooley:2008vu,Mediavilla:2009um}, pulsar timing 
arrays \cite{Bugaev:2010bb}, and FRB lensing searches \cite{Munoz:2016tmg} may also 
allow probes of the $\sim 30 \, M_\odot$ PBH mass range.

Another potential source of information is the stochastic GW background.
Models for the stochastic background due to BH mergers usually
entail a mass distribution that extends to smaller BH masses and
a redshift distribution that is somehow related to the
star-formation history.  Given microlensing limits, the PBH mass
function cannot extend much below $30\,M_\odot$.  Moreover, the
PBH merger rate per unit comoving volume is likely higher for
PBHs than for traditional BHs at high redshifts.  Together,
these suggest a stochastic background for PBHs that has more
weight at low frequencies and less at higher ones than
that from traditional BH sources.

The results of this work provide
additional motivation for more sensitive next-generation GW experiments 
such as the Einstein Telescope \cite{ET}, DECIGO \cite{Seto:2001qf} and BBO \cite{BBO}, which 
will continuously extend the aLIGO frequency range downwards. 
These may enable the tests described above for excesses in the BH mass spectrum, 
high ellipticity and low-frequency stochastic background that are required to determine if LIGO has detected dark matter.

\begin{acknowledgments}

We thank Liang Dai for useful discussions.  SB
was supported by NASA through Einstein Postdoctoral Fellowship
Award Number PF5-160133.  This work was supported by NSF Grant
No. 0244990, NASA NNX15AB18G, the John Templeton Foundation, and
the Simons Foundation.
\end{acknowledgments}

\end{document}